\documentclass[sigconf]{acmart}

\usepackage{booktabs} % For formal tables

\usepackage{balance}
\def\BibTeX{{\rm B\kern-.05em{\sc i\kern-.025em b}\kern-.08emT\kern-.1667em\lower.7ex\hbox{E}\kern-.125emX}}

\usepackage{subcaption}
\usepackage{float}

\newcommand{\hammingname}[0]{\ensuremath{d_H}}
\newcommand{\hamming}[2]{\ensuremath{\hammingname(#1,#2)}}
\newcommand{\hammingprojectedname}[0]{\ensuremath{\delta^P_H}}
\newcommand{\hammingprojected}[2]{\ensuremath{\hammingprojectedname(#1,#2)}}
\newcommand{\kl}[2]{\ensuremath{\textrm{KL}(#1||#2)}}
\newcommand{\norm}[1]{\ensuremath{\left\lVert#1\right\rVert}}

\newcommand{\proj}[2]{\ensuremath{\textrm{P}_{#1}(#2)}}

\copyrightyear{2021}
\acmYear{2021}
\setcopyright{iw3c2w3}
\acmConference[WWW '21]{Proceedings of the Web Conference 2021}{April 19--23, 2021}{Ljubljana, Slovenia}
\acmBooktitle{Proceedings of the Web Conference 2021 (WWW '21), April 19--23, 2021, Ljubljana, Slovenia}
\acmPrice{}
\acmDOI{10.1145/3442381.3450011}
\acmISBN{978-1-4503-8312-7/21/04}

\author{Christian Hansen}
\authornote{Both authors share the first authorship.}
\affiliation{
  \city{University of Copenhagen}
}
\email{chrh@di.ku.dk}

\author{Casper Hansen}
\authornotemark[1]
\affiliation{
  \city{University of Copenhagen}
}
\email{c.hansen@di.ku.dk}

\author{Jakob Grue Simonsen}
\affiliation{
  \city{University of Copenhagen}
}
\email{simonsen@di.ku.dk}

\author{Christina Lioma}
\affiliation{
  \city{University of Copenhagen}
}
\email{c.lioma@di.ku.dk}

\begin{document}
\fancyhead{}

\begin{abstract}
When reasoning about tasks that involve large amounts of data, a common approach is to represent data items as objects in the Hamming space where operations can be done efficiently and effectively. Object similarity can then be computed by learning binary representations (hash codes) of the objects and computing their Hamming distance. While this is highly efficient, each bit dimension is equally weighted, which means that potentially discriminative information of the data is lost. A more expressive alternative is to use real-valued vector representations and compute their inner product; this allows varying the weight of each dimension but is many magnitudes slower. To fix this, we derive a new way of measuring the dissimilarity between two objects in the Hamming space \textit{with} binary weighting of each dimension (i.e., disabling bits): we consider a field-agnostic dissimilarity that projects the vector of one object onto the vector of the other.
When working in the Hamming space, this results in a novel projected Hamming dissimilarity, which by choice of projection, effectively allows a binary importance weighting of the hash code of one object through the hash code of the other.
We propose a variational hashing model for learning hash codes optimized for this projected Hamming dissimilarity, and experimentally evaluate it in collaborative filtering experiments. The resultant hash codes lead to effectiveness gains of up to +7\% in NDCG and +14\% in MRR compared to state-of-the-art hashing-based collaborative filtering baselines, while requiring no additional storage and no computational overhead compared to using the Hamming distance.

%while significantly improving the converging rate during training compared to using the Hamming distance.

%Compared to the Hamming distance, our projected Hamming dissimilarity allows a bit-level importance coding without requiring any additional storage or computational overhead.

\end{abstract}


%\title{Unsupervised Pairwise Reconstruction based Semantic Hashing}
\title{Projected Hamming Dissimilarity for Bit-Level \\Importance Coding in Collaborative Filtering}
%\subtitle{WWW October 12/19 abstract/submission.}
\maketitle

\section{Introduction}
Hashing-based learning aims to find short binary representations of data objects (called \emph{hash codes}) that allow effective and efficient computations. For tasks involving multiple interacting objects, hash codes must be chosen carefully to ensure that certain properties of the codes (e.g., their inner products or mutual distance) carry information relevant to the task.
For example, in hashing-based collaborative filtering, binary user and item representations must be learned so that the distance between them reflects how much user $u$ likes item $i$. Currently, the most efficient way of doing this is to compute 
%Collaborative filtering \cite{herlocker1999algorithmic,Sarwar:2001:ICF:371920.372071} is the problem of predicting new items a user may like by processing preference information from other users, and is heavily used in modern personalized recommender systems.
% Early work on collaborative filtering is based on matrix factorization approaches \cite{Koren:2009:MFT:1608565.1608614} that learn a mapping to a shared $m$-dimensional real-valued space between users and items, such that user-item similarity can be estimated by the inner product. \textit{Hashing-based} collaborative filtering \cite{liu2014collaborative} extends this by allowing fast similarity searches that massively increase efficiency (e.g., real-time brute-force search in a billion items \cite{shan2018recurrent}). This is done by learning hash functions that map users and items into binary vector representations (\emph{hash codes}) and then using 
their Hamming distance, which is the sum of differing bits between two hash codes. However, by definition, the Hamming distance weighs each bit equally. This is a problem because the importance of the underlying properties encoded by each bit may differ.
An alternative is to use real-valued vectors and compute their inner product, which allows varying the weight of each dimension, and thus enables a dimension-specific importance weighting not possible using the Hamming distance for binary codes. 
However, binary codes allow large storage reductions compared to floating point representations, while the Hamming distance enables massively faster computations compared to the inner product (e.g., real-time brute-force search in a billion items \cite{shan2018recurrent}).
Motivated by this, we ask: can we introduce bit-level importance weighting on binary representations without compromising efficiency?

%In contrast, the inner product between real-valued vectors naturally enables importance weighting, but incurs efficiency costs because of the need for multiple floating point operations in high-dimensional spaces. However, 
We reason that, by the definition of the inner product, the distance between two vectors $u$ and $i$ should be identical to the difference in length between vector $u$ and vector $i$'s projection on $u$. This observation can be exploited by using a vector space where projections and lengths can be computed several magnitudes more efficiently than in Euclidean space. We show that performing the exact same projection and length computations in the Hamming vector space over $\mathbb{Z}_2$ results in a novel projected Hamming dissimilarity, which %\emph{naturally} 
corresponds to traditional measures used in real-valued representations. By choice of projection in the Hamming space, the projected Hamming dissimilarity effectively allows a bit-level binary weighting (corresponding to disabling bits) of $i$'s hash code via the hash code of $u$, but without decreasing efficiency compared to the Hamming distance. 
We propose a variational hashing model for learning hash codes optimized for our projected Hamming dissimilarity, which we experimentally evaluate in collaborative filtering experiments. 
Compared to state-of-the-art baselines using the Hamming distance, we observe effectiveness gains of up to +7\% in NDCG and +14\% in MRR, while also significantly improving the convergence rate during model training compared to optimizing with the Hamming distance.
 %We incorporate the projected Hamming dissimilarity into a new variational hashing-based collaborative filtering framework, called Projected Variational Hashing (PVH).%, which is designed for optimizing hash codes to be used with the projected Hamming dissimilarity. 
 
In summary, we \textbf{contribute} the first alternative to the Hamming distance for hashing-based learning that allows bit-level importance coding, while requiring no additional storage and no computational overhead.
We make our code publicly available.\footnote{The code is available at \url{https://github.com/casperhansen/Projected-Hamming-Dissimilarity}}

\section{Related Work}\label{s:relwork}
%We focus on collaborative filtering with \emph{explicit} feedback, which assumes that users and items are related via a rating specified by the user: the task is to rank a pool of pre-selected items. This is different from \emph{implicit} feedback, where the task is to estimate the pool of items that are of interest to the user.
Matrix factorization is one of the most popular collaborative filtering methods \cite{Koren:2009:MFT:1608565.1608614}, but to reduce storage requirements and speed up computation, hashing-based collaborative filtering has been researched. For hashing-based methods, the users and items are represented as binary hash codes (as opposed to real-valued vectors), which traditionally have used the highly efficient Hamming distance (as opposed to the inner product) for computing user-item similarities. In the following, we review the literature on both binary representation learning and importance coding of such binary representations. %\chh{Change to extremely efficient boolean operations?}
\subsection{Learning binary representations (hash codes)}
%\textbf{Two-stage approaches.} 
Early hashing-based collaborative filtering methods include two stages: First, real-valued representations of the data (vectors) are learned, and then the real-valued vectors are transformed into binary hash codes. \citet{Zhang:2014:PPH:2600428.2609578} use matrix factorization initially, followed by a binary quantization of rounding the real-valued vectors, while ensuring that the hash code is preference preserving with respect to observed properties of the data %of the observed user-item ratings 
using their Constant Feature Norm constraint. \citet{Zhou:2012:LBC:2339530.2339611} and \citet{liu2014collaborative} both explore binary quantization strategies based on orthogonal rotations of the real-valued vectors, which share similarities with Spectral Clustering \cite{Yu:2003:MSC:946247.946658}. However, the two-stage approaches often suffer from large quantization errors \cite{zhang2016discrete,Liu:2019:CCC:3331184.3331206}, because the hash codes are not learned directly, but rather based on different quantization procedures. More recently, hash codes are learned directly: this has been done using relaxed integer optimization while enforcing bit balancing and decorrelation constraints \cite{zhang2016discrete}; and (ii) using an autoencoder to learn the codes in an end-to-end manner \cite{hansen-coldstart-hash-2020}. The latter approach is most similar to the variational hashing model proposed in our work for optimizing hash codes for our projected Hamming dissimilarity, but their work is designed for cold-start recommendation based on generalizing hash codes from item descriptions. In contrast, our hashing model is designed to work based purely on user-item ratings, without assuming any additional knowledge of users or items.

%It is the latter approach that we choose to adapt for learning hash codes optimized for the projected Hamming dissimilarity, because of its high flexibility in modifying its objective function (see Section \ref{s:vh}).
%Extensions of DCF incorporate side-information (e.g., reviews associated with ratings) \cite{Lian:2017:DCM:3097983.3098008,Liu:2018:DFM:3304222.3304247,zhang2019neural} and have been redesigned for \emph{implicit} feedback signals \cite{zhang2017discrete}.

\subsection{Importance coding of binary representations} Approaches for coding importance into hash codes have been studied for the Hamming distance in several applications. \citet{zhang2013binary} present an image ranking approach that uses a weighted Hamming distance, where bit-level real-valued weights are learned based on both the discriminitative power of each bit across all hash codes, but also dynamically computed based on the hash code used for querying. The bit-level weights are multiplied on each differing bit between two hash codes, such that the distance is computed as the sum of the weights. Different ways of defining bit-level weights have been explored based on fixed weights per bit \cite{wang2013weighted}, fixed weights based on byte-level block differences between hash codes \cite{fan2013learning}, and query-adaptive weights \cite{ji2014query,zhang2018query}. While fixed weights enable faster retrieval than dynamic weights, they all share the same limitation of being significantly less efficient than the Hamming distance, because they can no longer be expressed using highly efficient Boolean hardware-level operations. Furthermore, in addition to the increased storage requirement due to the weights, transferring the weights to the lowest level of memory (i.e., the register) adds additional computational overhead compared to the Hamming distance.

More recent work addresses the problem that hash codes have reduced representational power compared to real-valued vectors, but increasing the hash code dimensionality to match the amount of bits used in the real-valued case hurts model generalization \cite{Liu:2019:CCC:3331184.3331206}. An alternative, in the task of collaborative filtering, is Compositional Coding for Collaborative Filtering (CCCF) \cite{Liu:2019:CCC:3331184.3331206}, which is a broadly similar method to learning compositional codes for (word) embedding compression \cite{chen2018learning,shu2018compressing}. CCCF is a hybrid of hash codes and real-valued weights: each hash code is split into $k$ blocks of $r$ bits each, and each block is associated with a real-valued scalar indicating the \textit{weight} of the block. The distance between two CCCF hash codes is then computed as a weighted sum of the Hamming distances of the individual blocks, where each weight is the product of each block's weight. The problem with this approach is that each block requires an individual Hamming distance computation, as well as floating point multiplications of the block weights. In fact, the CCCF block construction no longer allows for highly efficient Boolean operations because the distance computation is weighted by each block's weight. %Another problem is that it drastically increases storage requirements by needing to store the real-valued weights for all blocks in a hash code. 

In contrast to the %weighted Hamming approaches used 
above approaches, our projected Hamming dissimilarity can exploit the same highly efficient Boolean operations as the Hamming distance, while enabling a bit-level binary weighting on hash codes without reducing efficiency.

%our approach solves the above problems: it effectively allows to disable unimportant bits -- corresponding to a 1-bit block size with 0/1 weights -- without needing to store any additional weights or vectors. Additionally, our approach still only requires a single dissimilarity computation between the two hash codes. 

\section{Bit-level Importance Coding in Hash Codes}
%\jgs{On writing for NeuRIPS/ICML etc.: Be aware that the writing style is different from IR, so try to refrain from using pronouns such as "our" (which is an IR-thing). Write succinctly using the model name or write "the model". Keep "our approach" etc. to the intro.} Yep, we will do our best!

%Hashing-based learning aims at finding succinct binary representations (called \emph{hash codes}) that allow effective and efficient computation in subsequent learning tasks; for tasks involving multiple interacting objects, hash codes must be chosen carefully to ensure that certain properties of the codes--e.g., their inner products or mutual distance--carry information relevant to the learning task.
%For example, in hashing-based collaborative filtering, binary user and item representations are learned for a set of users and items, such that the distance between the representations indicates how well user $u$ likes item $i$. 
\subsection{Preliminaries} 
Given two data objects $u$ and $i$, let
$z_u \in \{-1,1\}^m$ and $z_i \in \{-1,1\}^m$ denote their hash codes, where $m$ is the number of bits in the hash code, which is typically chosen to fit into a machine word. 
%The preference of user $u$ for item $i$ is specified by the rating $R_{u,i} \in \{1,2,3,...,K\}$, where $K$ is the maximum rating, such that the distance between $z_u$ and $z_i$ is low when $R_{u,i}$ is high. 
The Hamming distance $\hammingname{}$ between $z_u$ and $z_i$ is defined as:
%Existing hashing-based collab%orative filtering approaches are optimized for the Hamming distance $\hammingname{}$ defined as:
\begin{align}\label{eq:hamming}
    \hamming{z_u}{z_i} = \sum_{j=1}^m  1_{ \big[ z_u^{(j)} \neq z_i^{(j)} \big]} = \text{SUM}(z_u \; \text{XOR} \; z_i)
\end{align}
which can be computed very efficiently due to the Boolean operations on the word level, and the SUM which is computed using the \emph{popcnt} bit string instruction. %Given a user and set of items, 
Because Hamming distance is integer-valued, the Hamming distances between several pairs of objects can be linear-time sorted using e.g. radix sort (Hamming distances must be elements of $[0,m]$, hence they are bounded) in ascending order to create a ranked list, allowing for very fast object similarity computation in data-heavy tasks like information retrieval \cite{shan2018recurrent}. 

The above definition clearly shows the efficiency strength of the Hamming distance, but also its weakness in 
%While the Hamming distance is highly efficient, it is by definition limited to 
weighting each bit-dimension equally, even though the underlying data properties encoded by some bits may be more discriminative than those of others. In contrast, representations using real-valued vectors and the inner product allow varying the weight of each dimension, and thus enable a dimension-specific importance weighting not possible using the Hamming distance for binary codes. Next, we show how we derive a new way of measuring the dissimilarity between $u$ and $i$ in the Hamming space \textit{with} binary weighting of each dimension. 

\subsection{Projected Hamming dissimilarity for bit-level importance coding} \label{ss:proj-ham-diss}
%To derive the dissimilarity with binary weighing in the Hamming space, we start by considering a general field agnostic dissimilarity. %, which in the Euclidean space corresponds to the well known cosine distance. This dissimilarity is then applied in the hamming space, to derive our proposed Hamming dissimilarity.
Let $V$ be a vector space over any field $F$, and let 
$\proj{\cdot}{\cdot}$ be a projection operator on $V$,
i.e., for each fixed $\vec{u}, \vec{i} \in V$,
$\proj{\vec{u}}{\cdot} : V \rightarrow V$ is a linear map such that
\begin{align}
\proj{\vec{u}}{\proj{\vec{u}}{\vec{i}}}) = \proj{\vec{u}}{\vec{i}}.
\end{align}
%$\innerproduct{\cdot}{\cdot} : V \times V  \longrightarrow F$ be an inner product on $V$. 
\noindent In the following, we consider an asymmetric relationship between two objects $\vec{u}$ and $\vec{i}$, such that $\vec{u}$ is considered a query object used for searching among data items denoted as $\vec{i}$. We consider both query and data items as elements of $V$. 
Intuitively, each dimension of V corresponds to a property of potential importance to a query (e.g., classical music); the projection of each query on the dimension corresponds to the strength of importance, and the projection of each item on the dimension corresponds to how much the item scores on that dimension.
%Intuitively, each dimension of $V$ corresponds to a property of the data, e.g., the specific color of a product in a recommender system setup.
%how expensive a product is in a recommender systems setup. %recommender systems potential importance to a query; 
%The projection of the query $\vec{u}$ unto item $\vec{i}$ means something different than the projection of the item $\vec{i}$ unto query $\vec{u}$: The projection of the query $\vec{u}$ unto item $\vec{i}$ corresponds to how important this property of the data (e.g., a specific color) is to the query (e.g., to the user of the recommender system). The projection of the item $\vec{i}$ to query $\vec{u}$ corresponds to how much of that property is represented in item $i$ (i.e. how expensive a product is).
%the projection of each query object on the dimension corresponds to the strength of importance, and the projection of each data object on the dimension corresponds to how much that object scores on that dimension.

Let $\norm{\cdot} : V \rightarrow \mathbb{R}$ be a norm on $V$; we define the \emph{dissimilarity}
between $\vec{u}$ and $\vec{i}$, denoted $\delta(\vec{u},\vec{i})$, as the norm of the projection
of $\vec{i}$ on $\vec{u}$:
\begin{align}\label{eq:delta}
\delta(\vec{u},\vec{i}) = \norm{\vec{u} - \proj{\vec{u}}{\vec{i}}}
\end{align}
%such that the dissimilarity is the length of the difference between the user vector and the projection of the item vector on the user vector. 
%Thus, the less the object agrees with the query needs, the larger the dissimilarity. 
Similarly to Eq. \ref{eq:hamming}, the more different $u$ and $i$ are, the higher their dissimilarity $\delta(\vec{u},\vec{i})$ should be. %%However, unlike the Hamming distance, $\delta(\vec{u},\vec{i})$ is not symmetric.
The dissimilarity is a natural concept: several existing notions of distance or similarity can be seen as special cases of Eq. \ref{eq:delta}. For example, in the standard Euclidean space the often-used cosine distance\footnote{We use the conventional naming of the cosine distance, even though it is not a proper distance as it does not satisfy the triangle inequality.}, $1-\cos(\vec{i},\vec{u})$, is one instance of Eq. \ref{eq:delta} if we assume unit length vectors. 

In hashing-based search, we are particularly interested in the binary vector space $V = \{-1,1\}^m$, with bitwise addition and scalar multiplication over the field $F = \mathbb{Z}_2$, where the projection operator is given by: 
\begin{align}
\proj{z_u}{z_i} = z_u \, \, \textrm{AND} \, \, z_i
\end{align}
\noindent for $z_u,z_i \in V$, i.e., masking the item hash code $z_i$ by the query hash code $z_u$. Due to working in the Hamming space, the norm $\norm{\cdot}$ is chosen as the Hamming norm (sometimes also called the zero norm or $L_0$). Using this, we obtain the projected Hamming dissimilarity $\hammingprojectedname{}$, defined as:
\begin{align}\label{eq:hamming-selfmask}
\hammingprojected{z_u}{z_i}= \norm{z_u - \proj{z_u}{z_i}} = \text{SUM}(z_u \, \, \textrm{XOR} \, \, \underbrace{(  z_u \, \, \text{AND} \, \, z_i )}_{\text{projection}} )
\end{align}
While having a similar formulation as the Hamming distance (see Eq. \ref{eq:hamming}), the projection of the item hash code $z_i$ unto the query hash code $z_u$ in Eq. \ref{eq:hamming-selfmask} allows a binary importance weighting of $z_i$, which corresponds to disabling unimportant bits as defined by the query hash code $z_u$ (corresponding to the bit-dimensions where the query hash code is -1). We consider bits to be disabled since a -1 bit in $z_u$ leads to \textit{all} item hash codes also having a -1 in that bit after the projection.
Note that due to the choice of projection, the projected Hamming dissimilarity is asymmetric (i.e., in general $\hammingprojected{z_u}{z_i} \neq \hammingprojected{z_i}{z_u}$), whereas the Hamming distance is symmetric.

% Interestingly, this derivation shows that the projected Hamming dissimilarity naturally corresponds to traditional dissimilarities in the Euclidean space. % #used in real-valued similarity search.

Compared to the Hamming distance, the projected Hamming dissimilarity fundamentally changes the purpose of the query hash code: instead of each dimension encoding a positive or negative preference for a property, it now encodes which properties of the item are important to the query (-1's from the query hash code are copied to the item due to the \texttt{AND} operation). 
Thus, this formulation can produce query-specific item representations while still only using a single code for each query and item respectively. 

\subsubsection{Speeding up the projected Hamming dissimilarity}
The projected Hamming dissimilarity in Eq. \ref{eq:hamming-selfmask} requires one additional Boolean operation compared to the Hamming distance. However, because the item codes are static once learned, we can reduce the time complexity to the same as the Hamming distance by observing that:
\begin{align}\label{eq:hamming-selfmask-fast}
\hammingprojected{z_u}{z_i} &= \text{SUM}(z_u \, \, \textrm{XOR} \, \, (  z_u \, \, \text{AND} \, \, z_i ) ) \nonumber \\
&= \text{SUM}(z_u \, \, \textrm{AND} \, \, (\textrm{NOT} \, \, z_i) )
\end{align}
where $(\textrm{NOT} \, \, z_i)$ can be precomputed and stored instead of the original item hash codes, thus requiring no additional storage and the same number of Boolean operations as the Hamming distance.

%The projected Hamming dissimilarity requires only one additional Boolean operation compared to the Hamming distance, but since this is applied once the codes are already placed in the lowest levels of memory (i.e., register), it only leads to a very marginal decrease in efficiency. %(see Section \ref{ss:run-time-analysis} for an empirical analysis).

Next, we present how the projected Hamming dissimilarity can be used for learning hash codes in collaborative filtering, where, notationally, a \textit{user} takes the role of the query, and \textit{items} are to be recommended based on their relevance to the user. 

\section{Projected Hamming Dissimilarity in Collaborative Filtering}\label{s:vh}
We propose a variational hashing model for collaborative filtering that learns user and item hash codes optimized for our projected Hamming dissimilarity. To derive a variational framework for hashing-based collaborative filtering, we define the likelihood of a user $u$ as the product over the likelihoods of the observed user specified ratings:
\begin{align}\label{eq:user_item_lkh1}
   p(u) = \prod_{i \in I_u} p(R_{u,i}), \;\;\;p(i) = \prod_{u \in U_i} p(R_{u,i}) 
\end{align}
where $I_u$ is the set of items rated by user $u$, and $U_i$ is the set of users who have rated item $i$. This formulation enforces a dual symmetric effect of users being defined by all their rated items, and items being defined by the ratings provided by all the users.
To maximize the likelihood of all observed items and users, we need to maximize the likelihood of the observed ratings $p(R_{u,i})$. Note that instead of maximizing the raw likelihood, we consider the log-likelihood to derive the objective below. We assume that the likelihood of a rating, $p(R_{u,i})$, is conditioned on two latent vectors: a user hash code $z_u$, and an item hash code $z_i$. We sample the hash code of a user or item by performing $m$ Bernoulli trials, which as a prior is assumed to have equal probability of sampling -1 and 1 ($p(z_i)$ and $p(z_u)$ below). This yields the following log-likelihood to be maximized:
\begin{align}\label{eq:user_item_lkh2}
%\log p_\theta(R_{u,i}) = \log \int\int_{\{-1,1\}^m} p_\theta(R_{u,i}|z_u,z_i) p(z_i)p(z_u) dz_u dz_i
\log p(R_{u,i}) =
\log \left( \sum_{\substack{z_i, z_u \in \{-1,1\}^{m}}} p(R_{u,i}|z_u,z_i) p(z_i)p(z_u) \right)
\end{align}
However, this is intractable to compute, hence we derive a lower bound. First, we define the learned approximate posterior distributions for the user and item hash codes as $q_\phi(z_i|i)$ and $q_\psi(z_u|u)$, respectively, where $\psi$ and $\phi$ are the distribution parameters. Next, we multiply and divide with the approximate posterior distributions, rewrite to an expectation, and finally apply Jensen's inequality to obtain a lower bound on the log-likelihood:
%The latent vectors $z_u$ and $z_i$ are user and item hash codes, and should therefore be conditioned on the user and item respectively. We denote the approximate posterior distributions for the hash codes as $q_\phi(z_i|i)$, $q_\psi(z_u|u)$, where $\psi$ and $\phi$ are the distribution parameters. Computing the log-likelihood directly is intractable, so we first multiple and divide with the approximate posterior distributions, from which we can rewrite to an expectation and apply Jensen's inequality to obtain a lower bound on the log-likelihood:
%To do this we first multiply and divide with the approximate posterior distributions $q_\phi(z_i|i)$, $q_\psi(z_u|u)$:
%\begin{align}
%\log p_\theta(R_{u,i}) = \log \sum_{\substack{z_i \in \{-1,1\}^{m} \\ z_u \in %\{-1,1\}^{m}}} p_\theta(R_{u,i}|z_u,z_i) p(z_i)p(z_u) %\frac{q_\phi(z_i|i)}{q_\phi(z_i|i)} \frac{q_\psi(z_u|u)}{q_\psi(z_u|u)}
%\end{align}
%where $\psi$ and $\phi$ are the parameters of the approximate posteriors. 
%We can now rewrite to an expectation and apply Jensen's inequality to obtain a lower bound on the log-likelihood:
\begin{align}
\log p(R_{u,i})
 \geq& \mathbb{E}_{q_\phi(z_i|i),q_\psi(z_u|u)} \Big[\log \big[ p(R_{u,i}|z_u,z_i) \big] \nonumber \\
 &+ \log p(z_i) - \log q_\phi(z_i|i) + \log p(z_u) - \log q_\psi(z_u|u)\Big] 
\end{align}
Since $z_i$ and $z_u$ will be sampled independently, then $q_\phi(z_i|i)$ and $q_\psi(z_u|u)$ are independent and we can rewrite to the variational lower bound:
\begin{align}
    \log p(R_{u,i}) \geq& \mathbb{E}_{q_\phi(z_i|i),q_\psi(z_u|u)} \Big[ \log \big[ p(R_{u,i}|z_u,z_i) \big] \Big] \nonumber \\ 
    & -\kl{q_\phi(z_i|i)}{p(z_i)} - \kl{q_\psi(z_u|u)}{p(z_u)}
\end{align}
where $\kl{\cdot}{\cdot}$ is the Kullback-Leibler divergence.
%\jgsinline{Don't we just have
%$$
%E_{q_\phi(z_i|i),q_\psi(z_u|u)} \Big[ \log [ %p_\theta(R_{u,i}|z_u,z_i) ] \Big] = \sum_{\substack{z_i \in \{-1,1\}^m\\
%z_u \in \{-1,1\}^m}}\log [ p_\theta(R_{u,i}|z_u,z_i) ] q_\phi(z_i|i)q_\psi(z_u|u)
%$$
%or am I missing something? Are the distributions $q_\phi(z_i|i)$ and $q_\psi(z_u|u)$ automatically independent? (Or do we assume they are?)}
Thus, to maximize the expected log-likelihood of the observed rating, we need to maximize the conditional log-likelihood of the rating, while minimising the KL divergence between the approximate posterior and prior distribution of the two latent vectors. Maximizing the expected conditional log-likelihood can be considered as a reconstruction term of the model, while the KL divergence can be considered as a regularizer.

Next we present the computation of the approximate posterior distributions $q_\phi(z_i|i)$ and $q_\psi(z_u|u)$ (Section \ref{s:post}) and the conditional log-likelihood of the rating $p(R_{u,i}|z_u,z_i)$ (Section \ref{s:cond_log}).
%$q^i_\phi(z_i|i)$, $q^u_\psi(z_u|u)$ and $p_\theta(R|z_u,z_i)$ are all approximated using a neural network.

%\begin{figure}
%    \centering
%    \includegraphics[width=0.48\textwidth]{fig/ijcai-smvh.pdf}
%    \vspace{-18pt}
%    \caption{SMVH model outline.
    %: The $m$-dimensional user and item sampling probabilities are learned and then sampled by repeating $m$ Bernoulli trials. The \texttt{AND} operation denotes the self-masking. The model is optimized using MSE between an affine transformation of the Hamming distance and the observed rating.
%    }
%    \label{fig:modeldesign}
%    \vspace{-5pt}
%\end{figure}

\subsection{Computing the approximate posterior distributions}\label{s:post}
The approximate posterior distributions can be seen as two encoder functions modelled as embedding layers in a neural network. Each encoder function maps either a user or an item to a hash code. We present below the derivation of the encoder function for the user (the item function is computed in the same way). The probability of the $j$'th bit is given by:
\begin{align}
    q^{(j)}_\psi(z_u|u) = \sigma(E^{(j)}_u)
\end{align}
where $E^{(j)}_u$ is the $j$'th entry in a learned real-valued embedding $E$ for user $u$, and $\sigma$ is the sigmoid function. The $j$'th bit is then given by:
\begin{align}\label{eq:hash-code-sampling}
    z_u^{(j)} = 2\lceil  \sigma(E^{(j)}_u) - \mu^{(j)} \rceil -1
\end{align}
where $\mu^{(j)}$ is either chosen stochastically by sampling $\mu^{(j)}$ from a uniform distribution in the interval [0,1], or chosen deterministically to be 0.5 (deterministic choice allows to obtain fixed hash codes for later evaluation). As the sampling is non-differentiable, a straight-through estimator \cite{bengio2013estimating} is used for backpropagation.

\subsection{Computing the conditional log-likelihood}\label{s:cond_log}
The conditional log-likelihood can be considered a reconstruction of the rating, given the user and item hash codes. Similarly to \cite{liang2018variational}, we model the observed ratings as a ground truth rating with additive Gaussian distributed noise, which is then discretized to the observed categorical rating. The conditional log-likelihood can then be computed as:
\begin{align}
    p(R_{u,i}|z_u,z_i) &= \mathcal{N}(R_{u,i}-f(z_u, z_i), \sigma^2)
\end{align}
where $f(z_u, z_i)$ is a function that reconstructs the rating given the user and item hash codes. Maximising the log-likelihood $\log  p(R_{u,i}|z_u,z_i) $ corresponds to minimising the mean squared error (MSE) between $R_{u,i}$ and $f(z_u, z_i)$, which is done for training the model. 
Existing work on hashing-based collaborative filtering \cite{Liu:2019:CCC:3331184.3331206,zhang2016discrete} also employs a MSE objective, and thus makes the same Gaussian distribution assumption as done here.

We define the reconstruction function to be our proposed projected Hamming dissimilarity:
\begin{align}
 f(z_u, z_i) = g(\hammingprojected{z_u}{z_i}) %g(\sum z_u^T z_i)   
\end{align}
where $g$ is a fixed affine transformation that maps the interval of the projected Hamming dissimilarity to the interval of the ratings, such that the minimum and maximum of the dissimilarity correspond to the minimum and maximum of the ratings. 
%The model is now fully differentiable and can be trained end-to-end using back propagation, such that the network is able to optimize the hash codes directly for self-masking \chh{Remove this sentence? add nothing (With this i mean "The model is now fully..."}. %A depiction of the model is provided in Figure \ref{fig:modeldesign}. 
It should be noted that while variational autoencoders are generative models, we do not explicitly utilize this in the model, as we are primarily concerned with the reconstruction of the observed ratings. This is standard in the related domain of semantic hashing \cite{shen2018nash,hansensemhash2019,hansen2020PairRec,hansen2021multiIndex}. 

\section{Experimental Evaluation}
We evaluate the effectiveness and efficiency of the projected Hamming dissimilarity for bit-level importance coding in hash codes in collaborative filtering experiments, where the task is to recommend relevant items to users. Items and users are represented as learned hash codes, and user-item relevance is approximated by operations (such as the Hamming distance or the projected Hamming dissimilarity) on those hash codes.

\subsection{Datasets} We use 4 publicly available datasets commonly used in prior work \cite{zhang2016discrete,Liu:2019:CCC:3331184.3331206,zhang2017discrete,Liu:2018:DFM:3304222.3304247,Lian:2017:DCM:3097983.3098008,rendle2009bpr} and summarized in the bottom of Table \ref{tab:mainresults}. The datasets comprise two movie rating datasets, Movielens 1M (ML-1M)\footnote{\url{https://grouplens.org/datasets/movielens/1m/}} and Movielens 10M (ML-10M)\footnote{\url{https://grouplens.org/datasets/movielens/10m/}}; a Yelp dataset with ratings of e.g., restaurant and shopping malls\footnote{\url{https://www.yelp.com/dataset/challenge}}; and a book rating dataset from Amazon\footnote{\url{http://jmcauley.ucsd.edu/data/amazon/}} \cite{he2016ups}. Following \citet{rendle2009bpr}, we remove users and items with less than 10 ratings. Following \citet{zhang2016discrete}, for each user 50\% of the ratings are used for testing, 42.5\% for training, and the last 7.5\% for validation. If a user rates an item multiple times, only the first rating is kept.

\subsection{Evaluation metrics}
\begin{table*}[t]
    \centering
        \caption{NDCG@k and MRR scores. $^*$ marks statistically significant gains over the other Hamming distance baselines per column using Bonferroni correction. $\Delta$\% shows the gain of VH$_{\text{PHD}}$ over the best hashing-based baseline per column.}
    \scalebox{0.93}{
    %\resizebox{\textwidth}{!}{
    \begin{tabular}{lccc|ccc|ccc|ccc}
     \toprule
        32 bit/dim. &  \multicolumn{3}{c|}{Yelp} &  \multicolumn{3}{|c|}{ML-1M} &  \multicolumn{3}{|c|}{ML-10M} &  \multicolumn{3}{|c}{Amazon} \\
         &  \tiny NDCG@5 & \tiny NDCG@10 & \tiny MRR
         &  \tiny NDCG@5 & \tiny NDCG@10 & \tiny MRR
         &  \tiny NDCG@5 & \tiny NDCG@10 & \tiny MRR
         &  \tiny NDCG@5 & \tiny NDCG@10 & \tiny MRR \\ \midrule
\multicolumn{2}{l}{\textbf{Hamming distance}} && &&& &&& &&& \\
CCCF & .7286 & .8000 & .6250 & .6867 & .7110 & .6493 & .5491 & .5987 & .5683 & - & - & - \\
DCF & .7412 & .8095 & .6368 & .6791 & .7092 & .6382 & .5645 & .6120 & .5843 & .8256 & .8714 & .7759 \\ 
MFmean & .6912 & .7712 & .5815 & .5631 & .5950 & .5053 & .4111 & .4688 & .4271 & .7899 & .8452 & .7342 \\
MFmedian & .6935 & .7734 & .5769 & .5631 & .5952 & .5085 & .4082 & .4665 & .4225 & .7899 & .8452 & .7343 \\
VH  & .7467 & .8132 & .6473 & .6851 & .7123 & .6419 & .5669 & .6157 & .5815 & .8254 & .8712 & .7758 \\ 
\midrule
\multicolumn{3}{l}{\textbf{Projected Hamming dissimilarity}} & &&& &&& &&& \\
VH$_{\textrm{PHD}}$ & \textbf{.8036}$^*$ & \textbf{.8547}$^*$ & \textbf{.7406}$^*$ & \textbf{.7135}$^*$ & \textbf{.7360}$^*$ & \textbf{.6940}$^*$ & \textbf{.5939}$^*$ & \textbf{.6358}$^*$ & \textbf{.6235}$^*$ & \textbf{.8479}$^*$ & \textbf{.8877}$^*$ & \textbf{.8062}$^*$ \\
%$\Delta$\% & \textbf{+8.4\%} & \textbf{+5.6\%} & \textbf{+16.3\%} & \textbf{+3.9\%} & \textbf{+3.5\%} & \textbf{+6.9\%} & \textbf{+5.2\%} & \textbf{+3.9\%} & \textbf{+6.7\%} & \textbf{+2.7\%} & \textbf{+1.9\%} & \textbf{+3.9\%} \\ \midrule

$\Delta$\% & \textbf{+7.6\%} & \textbf{+5.1\%} & \textbf{+14.4\%} & \textbf{+3.9\%} & \textbf{+3.3\%} & \textbf{+6.9\%} & \textbf{+4.8\%} & \textbf{+3.3\%} & \textbf{+6.7\%} & \textbf{+2.7\%} & \textbf{+1.9\%} & \textbf{+3.9\%} \\ \midrule

\textbf{Inner product} &&& &&& &&& &&& \\
MF & .8071$^*$ & .8573$^*$ & .7513$^*$ & .7352$^*$ & .7502$^*$ & .7370$^*$ & .6029$^*$ & .6427$^*$ & .6385$^*$ & .8586$^*$ & .8954$^*$ & .8262$^*$ \\\bottomrule\bottomrule %\vspace{2pt}
    \end{tabular}}
    %\vspace{15pt}
    \scalebox{0.93}{
    %\resizebox{\textwidth}{!}{
    \begin{tabular}{lccc|ccc|ccc|ccc}
     
        %64 bits &  \multicolumn{3}{c|}{Yelp} &  \multicolumn{3}{|c|}{ML-1M} &  \multicolumn{3}{|c|}{ML-10M} &  \multicolumn{3}{|c|}{Amazon} \\
        64 bit/dim. &  \tiny NDCG@5 & \tiny NDCG@10 & \tiny MRR
         &  \tiny NDCG@5 & \tiny NDCG@10 & \tiny MRR
         &  \tiny NDCG@5 & \tiny NDCG@10 & \tiny MRR
         &  \tiny NDCG@5 & \tiny NDCG@10 & \tiny MRR \\ \midrule
         \multicolumn{2}{l}{\textbf{Hamming distance}} && &&& &&& &&& \\
CCCF & .7371 & .8060 & .6329 & .7016 & .7259 & .6716 & .5645 & .6134 & .5837 & - & - & - \\
DCF & .7497 & .8155 & .6574 & .7049 & .7285 & .6766 & .5865 & .6316 & .6088 & .8299 & .8747 & .7825 \\ 
MFmean & .6912 & .7712 & .5810 & .5666 & .5981 & .5172 & .4104 & .4675 & .4257 & .7902 & .8458 & .7340 \\
MFmedian & .6954 & .7752 & .5780 & .5649 & .5966 & .5105 & .4113 & .4679 & .4270 & .7902 & .8457 & .7334 \\
VH  & .7537 & .8185 & .6561 & .7103 & .7338 & .6759 & .5860 & .6328 & .6013 & .8300 & .8746 & .7828 \\ 
\midrule 
\multicolumn{3}{l}{\textbf{Projected Hamming dissimilarity}} & &&& &&& &&& \\
VH$_{\textrm{PHD}}$  & \textbf{.8075}$^*$ & \textbf{.8577}$^*$ & \textbf{.7540}$^*$ & \textbf{.7267}$^*$ & \textbf{.7459}$^*$ & \textbf{.7136}$^*$ & \textbf{.6034}$^*$ & \textbf{.6427}$^*$ & \textbf{.6373}$^*$ & \textbf{.8521}$^*$ & \textbf{.8908}$^*$ & \textbf{.8147}$^*$ \\
%$\Delta$\% & \textbf{+7.7\%} & \textbf{+5.2\%} & \textbf{+14.7\%} & \textbf{+3.1\%} & \textbf{+2.4\%} & \textbf{+5.5\%} & \textbf{+2.9\%} & \textbf{+1.8\%} & \textbf{+4.7\%} & \textbf{+2.7\%} & \textbf{+1.8\%} & \textbf{+4.1\%} \\ \midrule

$\Delta$\% & \textbf{+7.1\%} & \textbf{+4.8\%} & \textbf{+14.7\%} & \textbf{+2.3\%} & \textbf{+1.7\%} & \textbf{+5.5\%} & \textbf{+2.9\%} & \textbf{+1.6\%} & \textbf{+4.7\%} & \textbf{+2.7\%} & \textbf{+1.8\%} & \textbf{+4.1\%} \\ \midrule

\textbf{Inner product} &&& &&& &&& &&& \\
MF & .8096$^*$ & .8591$^*$ & .7573$^*$ & .7424$^*$ & .7552$^*$ & .7439$^*$ & .6188$^*$ & .6562$^*$ & .6594$^*$ & .8586$^*$ & .8954$^*$ & .8259$^*$ \\

 \bottomrule 
 \toprule %\vspace{2pt}
    \textbf{Dataset} & \multicolumn{3}{l|}{22,087 users} & \multicolumn{3}{l|}{6,040 users} & \multicolumn{3}{l|}{69,878 users} & \multicolumn{3}{l}{158,650 users} \\
    
    \textbf{properties} & \multicolumn{3}{l|}{14,873 items} & \multicolumn{3}{l|}{3,260 items} & \multicolumn{3}{l|}{9,708 items} & \multicolumn{3}{l}{128,939 items} \\
    
     & \multicolumn{3}{l|}{602,517 ratings} & \multicolumn{3}{l|}{998,539 ratings} & \multicolumn{3}{l|}{9,995,471 ratings} & \multicolumn{3}{l}{4,701,968 ratings} \\
    
    & \multicolumn{3}{l|}{0.18\% density} & \multicolumn{3}{l|}{5.07\% density} & \multicolumn{3}{l|}{1.47\% density} & \multicolumn{3}{l}{0.02\% density} \\
    
    \bottomrule

    \end{tabular}}
    \label{tab:mainresults}
\end{table*}
We measure the effectiveness with the mean Normalised Discounted Cumulative Gain (NDCG) \cite{jarvelin2000ir}, which is often used to evaluate recommender systems with non-binary ratings (or relevance values). We use the average NDCG at cutoffs $\{5,10\}$ over all users and report the average for each cutoff value. We also compute the Reciprocal Rank (RR) of the highest rated item per user, averaged over all users, which represents how well the approaches are at getting a highly relevant item to the top of the ranked list:
\begin{align}
     \text{DCG@k} = \sum_{i=1}^k \frac{2^{\text{rel}_i}-1}{\log_2(i+1)}, \;
     \text{NDCG@k} = \frac{\text{DCG@k}}{\text{DCG@k}^{(\text{opt})}}, \;
     \text{RR} = \frac{1}{\text{rank}}
 \end{align}
where rel$_i$ is the relevance of the item in position $i$ of the ranked list of items, $\text{DCG@k}^{(\text{opt})}$ is the DCG@k of the optimal ranking, and rank is the position of the first highest rated item for a user. These measures are averaged across all users, and following the standard notation, the mean reciprocal rank is denoted as MRR.

\subsection{Baselines}\label{ss:baselines}
We learn hash codes optimized for the projected Hamming dissimilarity by incorporating it in the variational hashing model as described in Section \ref{s:vh} (denoted VH$_{\textrm{PHD}}$). We compare this to standard and state-of-the-art baselines (listed below) that use Hamming distance in different ways. 
%, but none of them allow efficient bit-level importance coding. 
For reference, we also include real-valued Matrix Factorization (MF)\footnote{Included as baseline in the CCCF repository \url{https://github.com/3140102441/CCCF}} \cite{Koren:2009:MFT:1608565.1608614}. MF is not directly comparable to the hashing approaches (it learns latent real-valued vectors for users and items, instead of hash codes, and computes their inner product), but we include it as a point of reference to a real-valued collaborative filtering baseline. In fact, MF has been shown to be both more efficient and highly competitive in effectiveness compared to neural real-valued collaborative filtering approaches \cite{rendle2020neural}. For a fair comparison of MF to the hashing approaches, we set the latent dimension in MF to be the same as the number of bits used in the hashing approaches. We experiment with hash code lengths of 32 and 64 bits, which correspond to the common machine word sizes. The hashing baselines are: 

%We use as baselines the state-of-the-art methods for hashing-based collaborative filtering (see Section \ref{s:relwork}), (real-valued) matrix factorisation, and two different strategies for binarizing the output of the matrix factorisation. Similarly to \cite{zhang2016discrete,Liu:2019:CCC:3331184.3331206}, we include matrix factorization \cite{Koren:2009:MFT:1608565.1608614} as a reference to a real-valued collaborative filtering baseline. In fact, matrix factorization has been shown to be both more efficient and highly competitive in effectiveness compared to neural real-valued collaborative filtering approaches \cite{rendle2020neural}, and thus serve as a good baseline to highlight the performance gap between the hashing-based and real-valued approaches.

\begin{itemize}
    \item \textbf{DCF}\footnote{\url{https://github.com/hanwangzhang/Discrete-Collaborative-Filtering}} \cite{zhang2016discrete} learns user and item hash codes through binary matrix factorization solved as a relaxed integer problem, while enforcing bit balancing and decorrelation constraints. The codes can be used directly for computing the Hamming distance.
    \item \textbf{CCCF}\footnote{\url{https://github.com/3140102441/CCCF}} \cite{Liu:2019:CCC:3331184.3331206} learns hash codes of $k$ blocks, where each block has $r$ bits. A floating point weight is given to each block for computing user-item similarities as a weighted sum of block-level Hamming distances. In \cite{Liu:2019:CCC:3331184.3331206}, the floating point weights are mistakenly not counted towards the amount of bits used, thus leading to an unfair advantage. For a fair comparison, we count each floating point weight only as 16 bits (rather than the typical 32 or 64 bits used for single or double precision, respectively). 
   % \item \textbf{MF}\footnote{Provided as a baseline in the CCCF repository  \url{https://github.com/3140102441/CCCF}} \cite{Koren:2009:MFT:1608565.1608614} is the classical matrix factorization based collaborative filtering method, where latent real-valued vectors are learned for users and items. We set the latent dimension to be the same as the number of bits used in the hashing-based approaches.
    \item \textbf{MFmean} and \textbf{MFmedian} are based on matrix factorization (MF), but use each dimension's mean or median for doing the binary quantization to bits \cite{Zhang:2010:SHF:1835449.1835455}. Similar to DCF, these codes can be used directly for computing the Hamming distance. We include these to highlight the large quantization loss occurring when the hash codes are not learned directly.
    \item \textbf{VH} is the same variational hashing model that we use for learning hash codes to be used with the projected Hamming dissimilarity, but here the codes are learned using the Hamming distance.
\end{itemize}

%\subsection{Tuning}
\subsection{Tuning} All hyper parameters for the baselines are tuned using the same set of possible values as in the original papers. For CCCF, we use block sizes of $\{8,16,32,64\}$ and each floating point weight counts for 16 bits. We try all possible combinations that fit within the bit budget, and if a single block is chosen, then the weight is not included in the bit calculation. Using a Titan X GPU, we train our variational hashing model from Section \ref{s:vh} using the Adam optimizer \cite{kingma2014adam}, and tune the learning rate from the set $\{0.005,\textbf{0.001},0.0005\}$ and the batch size from the set $\{100,200,\textbf{400},800\}$ (best values in bold). As noise injection has been found beneficial to reduce over-fitting in variational neural models \cite{sohn2015learning}, we add Gaussian noise to the ratings during training; we initially set the variance of the Gaussian to 1 and reduce by a factor of $1-10^{-4}$ in every training iteration.

\subsection{Effectiveness results}\label{ss:results}

Table \ref{tab:mainresults} reports the effectiveness results measured with NDCG and MRR, where the highest NDCG and MRR per column among the hashing-based baselines is shown in \textbf{bold}. Results statistically significantly better than the other Hamming distance baselines per column, using a paired two tailed t-test at the 0.05 level and Bonferroni correction, are indicated by an asterisk $^*$. The Amazon results for CCCF are not included because the released implementation requires $>$128GB of RAM on this dataset due to the larger amount of items and users. %\cl{biggest gains in yelp. Why? -- cannot find any obvious reason}

There are 4 main findings in Table \ref{tab:mainresults}: (1) Hash codes optimized for the projected Hamming distance (VH$_{\textrm{PHD}}$) outperform all hashing baselines at all times. (2) The gains of VH$_{\textrm{PHD}}$ are larger for MRR than for NDCG, which means that the bit-level importance coding impacts the very top of the ranking list (i.e., the recommendations that matter the most). (3) The best hashing baselines (CCCF, DCF, and VH) have overall similar scores, which indicates a potential ceiling in effectiveness with standard Hamming distance on hash codes. (4) MF with real-valued vectors (i.e., no hash codes) using the inner product outperforms all the hashing approaches, which is to be expected as the representational power of 32/64 floating point numbers is notably higher than that of 32/64 bits. However, VH$_{\textrm{PHD}}$ bridges a large part of the effectiveness gap between the hashing baselines and MF, such that the NDCG differences between VH$_{\textrm{PHD}}$ and MF in 9 out of 16 cases are below 0.01, while the MRR differences in 4 out of 8 cases are close to 0.01.

%\subsubsection{Improvements across users}\label{sss:improv-across-users}
\subsubsection{Impact of user difficulty}\label{sss:improv-across-users}
\begin{figure*}
    \centering
    \includegraphics[width=0.95\textwidth]{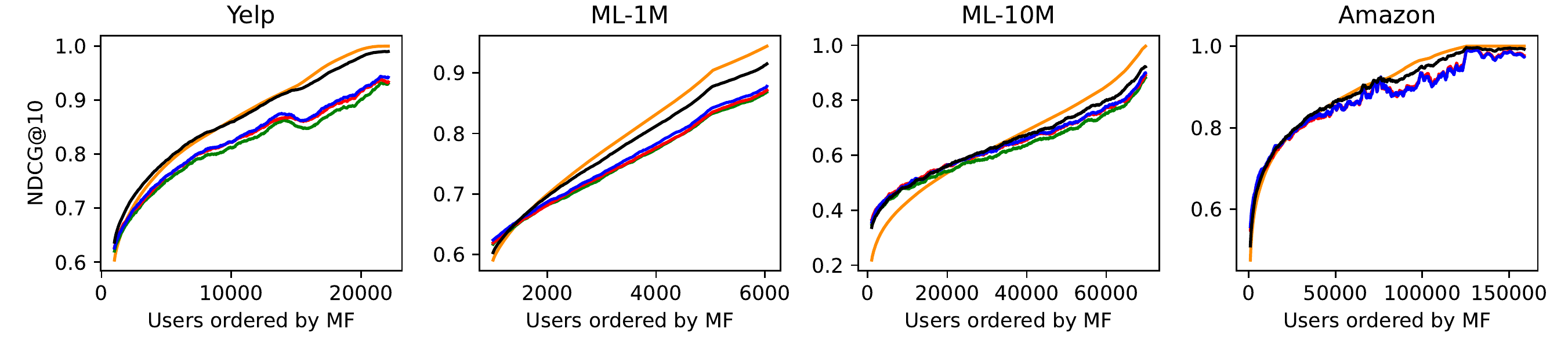}
    \vspace{-8pt}
    \caption{Users are ordered by NDCG@10 for MF and the user-level performances are plotted.}
    \label{fig:perf-comp}
    \vspace{-10pt}
\end{figure*}

Given MF as the best performing method, we consider each user's MF performance to be an indicator of difficulty for modeling that particular user. 
To see how this type of user difficulty impacts recommendation performance, we sort all users (per dataset) increasingly according to their 64-dimensional MF NDCG@10 scores (x axis), and plot the average NDCG@10 score per user smoothed by averaging the 500 nearest users (y axis). We do this for the three best Hamming distance baselines (CCCF, DCF, and VH), VH$_{\text{PHD}}$, and MF, which can be seen in Figure \ref{fig:perf-comp}. 

We observe that VH$_{\text{PHD}}$ outperforms all Hamming distance baselines, showing that the projected Hamming dissimilarity is robust across users. Note that, for the 20,000 users with the lowest MF NDCG@10 on ML-10M, all hashing-based methods outperform MF, highlighting that MF is not always consistently better than hashing-based alternatives (despite allowing for much richer (real-valued versus binary) data representations). In addition, for Yelp and Amazon, VH$_{\textrm{PHD}}$ obtains near-identical performance as MF for a majority of the users. Interestingly, on Amazon the hashing-based approaches generally perform worse than MF on the $[80000, 120000]$ user interval. We argue that this observation is due to those users not being expressed well by the limited representational power of hash codes using the Hamming distance, compared to real-valued vectors using the inner product, but the projected Hamming dissimilarity reduces a large part of this gap.

\begin{figure*}
    \centering
    \includegraphics[width=0.95\textwidth]{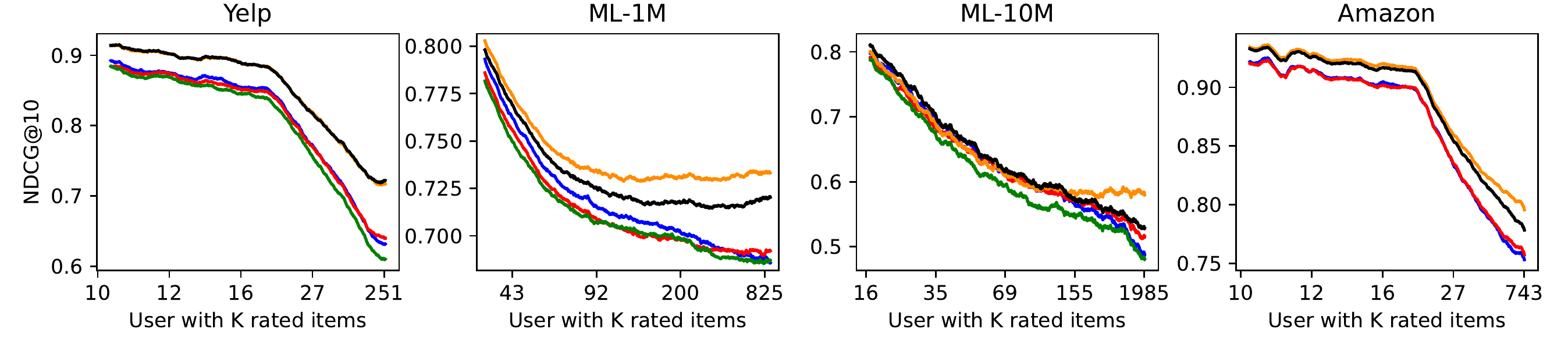}
    \vspace{-8pt}
    \caption{Users are ordered by their number of rated items and the user-level NDCG@10 are plotted.}
    \label{fig:k-rated-comp}
    \vspace{-10pt}
\end{figure*}
\begin{figure*}
    \centering
    \includegraphics[width=0.95\textwidth]{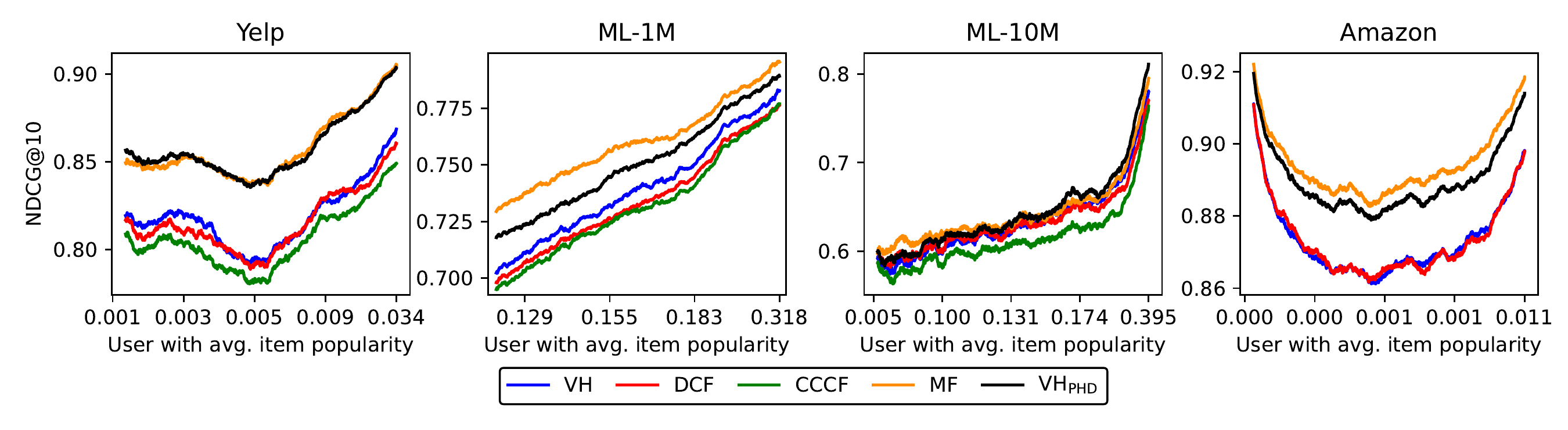}
    \vspace{-12pt}
    \caption{Users are ordered by their average item popularity and the user-level NDCG@10 are plotted.}
    \label{fig:pop-comp}
    \vspace{-5pt}
\end{figure*}

\subsubsection{Impact of the number of items per user} \label{sss:impact-num-users}
%In this section and the next, we investigate how two different user characteristics impact the performance of the methods. First, in Figure \ref{fig:k-rated-comp} 
We investigate the impact of a user's activity as defined by their number of rated items and plot the NDCG@10 scores per user for 64 bit hash codes and 64-dimensional vectors for MF (See Figure \ref{fig:k-rated-comp}). 

Generally for all methods, we observe higher performance for users with few rated items, and the performance drops as the number of rated items increases. On all datasets except Yelp, the hashing-based baselines perform similar to the real-valued MF initially, but the performance gap occurs as the number of rated items increases, especially for the users with the highest number of rated items for ML-10M and Amazon. The hash codes may perform worse on users with a high number of rated items due to their limited representational power (compared to a real-valued vector), however, using the projected Hamming dissimilarity in our VH$_{\textrm{PHD}}$ enables to reduce this gap significantly. In fact, our VH$_{\textrm{PHD}}$ performs almost identically to MF on Yelp and most of the users on ML-10M and Amazon, except those with the highest number of rated items.

\subsubsection{Impact of the average item popularity per user}
We now consider how a user's average item popularity impacts performance. We denote an item's popularity as the fraction of users who have rated the item, such that a user's average item popularity can vary between 0.0 to 1.0, where 1.0 corresponds to only having rated items all other users have rated as well. We order users by their average item popularity and plot the NDCG@10 scores per user for 64 bit hash codes and 64-dimensional vectors for MF (See Figure \ref{fig:pop-comp}).

Generally for all methods, users with a high average item popularity obtain the highest performance scores, whereas users with a lower average item popularity tend to be harder to model. This can be explained by popular items appearing often during training, thus making it easier to learn high quality item representations (both binary and real-valued) matching the properties of those items. Interestingly, on Amazon and partly Yelp, the performance increases for users with the lowest average item popularity. We argue this is primarily due to the high sparsity of those datasets, meaning that some items are rated by very few users, thus making it possible to fit to a small set of user types. Similarly to the analysis on the number of rated items in Section \ref{sss:impact-num-users}, we overall observe a similar trend that our VH$_{\textrm{PHD}}$ significantly reduces the gap between the existing hashing-based methods and the real-valued MF.

%and we generally observe the smallest difference between VH$_{\textrm{PHD}}$ and MF in these cases.

\subsubsection{Stochastic or deterministic hash codes} 
\begin{figure}
    \centering
     \includegraphics[width=0.87\linewidth]{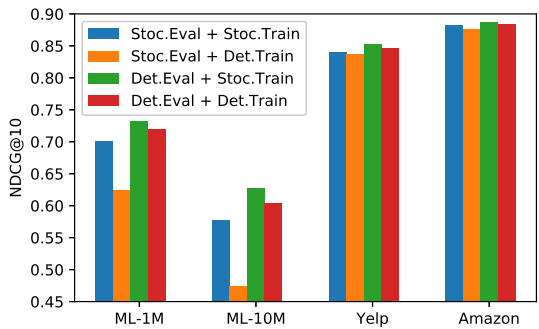}
     \vspace{-8pt}
     \caption{NDCG@10 of VH$_{\textrm{PHD}}$ when varying whether 64 bit hash codes are sampled stochastically or deterministically.}
     \label{fig:analyse_stoc_det}
     \vspace{-5pt}
\end{figure}

We investigate the effect of the sampling strategy for hash codes (see Eq. \ref{eq:hash-code-sampling}) during training and evaluation. 
The sampling can either be deterministic ($\mu^{(j)} = 0.5$) or stochastic ($\mu^{(j)}$ is sampled uniformly at random from $[0,1]$), and does not have to be the same for training and evaluation. 
Figure \ref{fig:analyse_stoc_det} shows the performance for the four configurations of stochastic sampling or deterministic output across all datasets. We observe that stochastic training with deterministic evaluation consistently performs the best, while deterministic training and deterministic evaluation perform second best. As expected, stochastic sampling at evaluation performs significantly worse than the deterministic option (even more so when trained deterministically), as every item has a small probability of being sampled such that it has a small distance to a user, even though it has a low rating (and vice versa for highly rated items). 

\begin{figure*}
    \centering
    \includegraphics[width=0.245\textwidth]{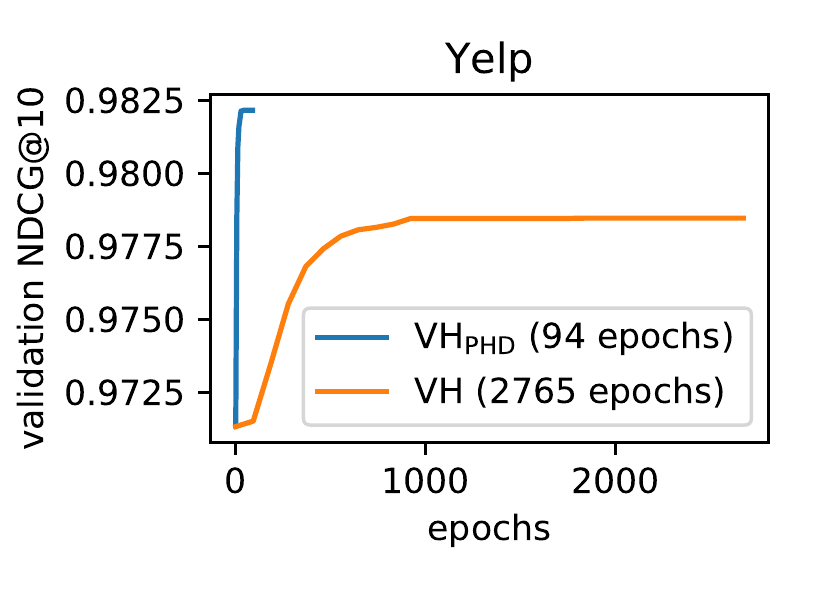}\hspace{-7pt}
    \includegraphics[width=0.245\textwidth]{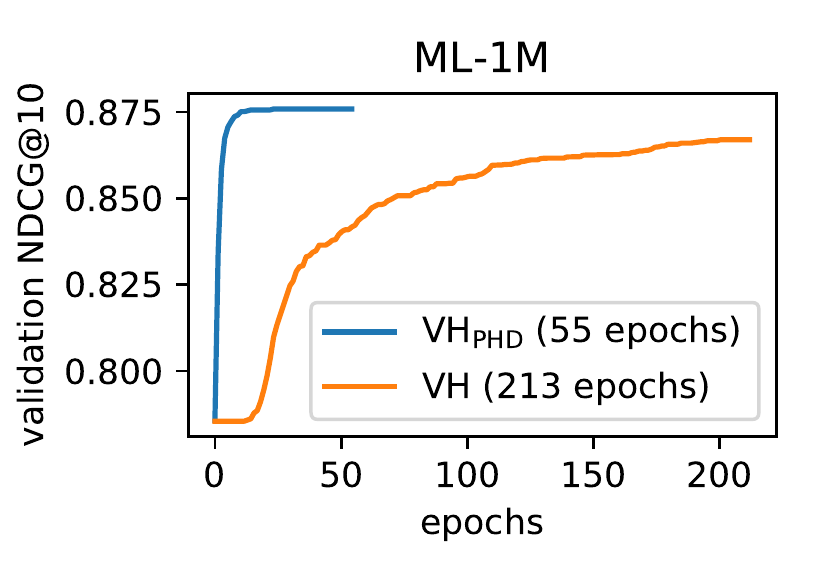}\hspace{-7pt}
    \includegraphics[width=0.245\textwidth]{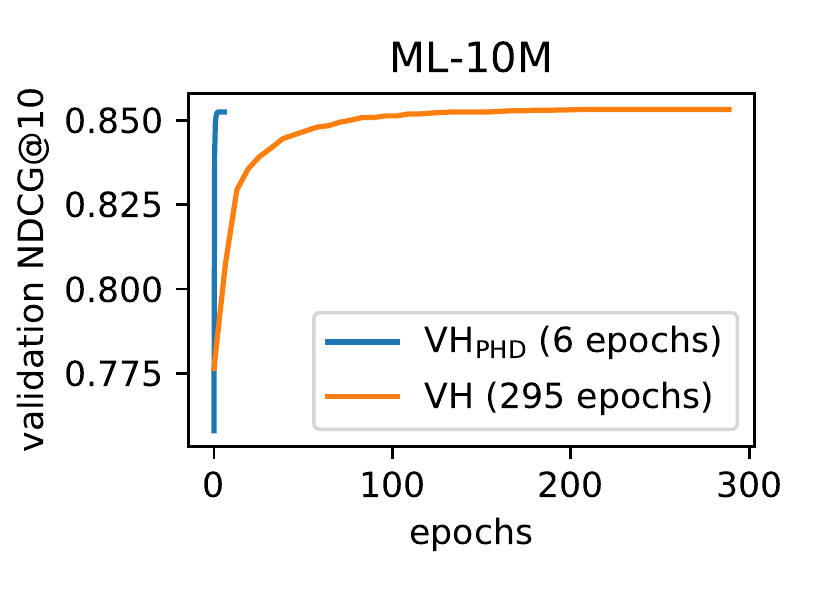}\hspace{-7pt}
    \includegraphics[width=0.245\textwidth]{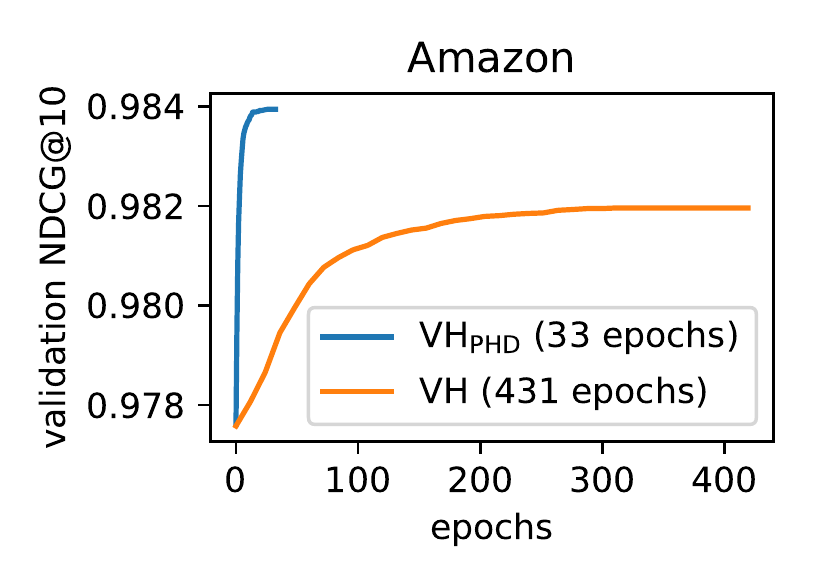}
    \vspace{-16pt}
    \caption{Convergence rate optimizing projected Hamming dissimilarity or Hamming distance.}
    \label{fig:analyseconv1}
    \vspace{-5pt}
\end{figure*}

\subsection{Efficiency results}
We measure the efficiency of the projected Hamming dissimilarity in terms of convergence rate (when integrated into the variational hashing model) and runtime overhead compared to the Hamming distance.

%\textbf{How the projected Hamming dissimilarity influences the convergence rate of the model.} 

\subsubsection{Convergence rate}
Figure \ref{fig:analyseconv1} shows the convergence rate for the variational hashing model using either the Hamming distance or the projected Hamming dissimilarity for 64 bit hash codes. We see that training with the projected Hamming dissimilarity significantly improves the convergence rate compared to the model with the Hamming distance. The time to run a single batch is the same for both the Hamming distance and the projected Hamming dissimilarity (approximately 0.007 seconds for a batch size of 400), from which we conclude that using and optimizing for the projected Hamming dissimilarity not only improves NDCG and MRR, but also drastically reduces training time. 
We argue that the masking done in the projected Hamming dissimilarity makes the hash codes easier to learn: During training, any update in the item hash code that makes a bit flip will change the distance to all users. However, for the projected Hamming dissimilarity, an update to an item only influences the projected Hamming dissimilarity to a user if the user's corresponding bit is 1 (as opposed to -1). Thus, the Hamming distance has a global reach for each bit, compared to a more localised reach for the projected Hamming dissimilarity, which effectively makes the hash codes easier to learn.

% We argue that this is due to the effective bit space being notably reduced by the projection within the projected Hamming dissimilarity, since all dimensions masked-out by the projection are effectively ignored. During training, this is beneficial because it allows the model to make local bit flips that only affect a subset of users (or items), whereas with the Hamming distance a bit flip always has a global reach, thus impacting all users (or items).

% \begin{figure*}[]
%   \begin{minipage}{\textwidth}
%   \begin{minipage}[b]{0.49\textwidth}
  
%     \includegraphics[width=0.495\textwidth]{fig/conv-yelp.eps}
%     \includegraphics[width=0.495\textwidth]{fig/conv-amazon.eps} 
%     \includegraphics[width=0.495\textwidth]{fig/conv-ml1m.eps} 
%     \includegraphics[width=0.495\textwidth]{fig/conv-ml10m.eps}
    
%     \caption{Convergence rate using the validation NDCG@10.}
%     \label{fig:analyseconv1}
    
%   \end{minipage}
%   \hfill
%   \begin{minipage}[b]{0.49\textwidth}
    
%         \centering
%     \includegraphics[width=1\textwidth]{fig/stoc_det}
%     \caption{NDCG@10 of \ourmethod\ when varying whether hash codes are sampled stochastically or deterministically.}
%     \label{fig:analyse_stoc_det}
%     \vspace{-3pt}
%     \end{minipage}
%   \end{minipage}
%   \vspace{-1pt}
% \end{figure*}

%\subsection{Runtime Analysis}\label{ss:run-time-analysis}

\subsubsection{Runtime analysis}
The projected Hamming dissimilarity has the same number of Boolean operations as the Hamming distance (see Eq. \ref{eq:hamming} and \ref{eq:hamming-selfmask-fast}), due to exploiting that the negation of the item hash codes in the projected Hamming dissimilarity can be precomputed and stored instead of the original item hash codes (i.e., requiring no additional storage).
We now verify the actual runtime of both the Hamming distance and the projected Hamming dissimilarity, when considering a fixed set of hash codes. Note that this is a comparison of generated hash codes, thus all approaches using the Hamming distance (or projected Hamming dissimilarity) would have the same runtime in the following experiment.
We implement both the Hamming distance and projected Hamming dissimilarity efficiently in C on a machine with a 64 bit instruction set. A test environment was made with 100M randomized 64 bit hash codes, where we measure the time taken to compute 100M Hamming distances and projected Hamming dissimilarities (averaged over 1000 repeated runs).
All experiments were run on a single thread\footnote{We used a Intel Core i9-9940X @ 3.30GHz and had 128GB RAM available.}, with all hash codes loaded in RAM. The source code was compiled with the highest optimization level utilizing all optimization flags applicable to the hardware.

As reported in Table \ref{tab:runtime}, the mean experiment time was 0.07401 seconds using both the Hamming distance the projected Hamming dissimilarity. Thus, the projected Hamming dissimilarity add no computational overhead, but still allows learning hash codes enabling significant effectiveness improvements as reported in Table \ref{tab:mainresults}. Finally, Table \ref{tab:runtime} also reports the runtime of computing the inner product of floating point vectors of length 64: the computation time is 4.71414 seconds, thus being significantly slower than the Hamming space operations.

\begin{table}
    \centering
    \caption{Runtime in seconds and runtime overhead compared to the Hamming distance for 100M computations.}
    %\vspace{-3pt}
    \resizebox{\linewidth}{!}{
    \begin{tabular}{lrr}
    \toprule
         & Runtime (s) & Runtime overhead \\
         \midrule
        Hamming distance & 0.07401 & - \\
        Projected Hamming dissimilarity  & 0.07401 & +0.0\% \\
        Inner product & 4.71414 &  +6269.6\%\\
        \bottomrule
    \end{tabular}}
    \label{tab:runtime}
    %\vspace{-5pt}
\end{table}

\section{Conclusion}
We presented the projected Hamming dissimilarity, which allows bit-level binary importance weighting (i.e., disabling bits), to produce hash codes that accurately represent dissimilarity between data objects and allow for very efficient subsequent processing. Next, we proposed a variational hashing model for learning hash codes to be optimized for the projected Hamming dissimilarity, and experimentally evaluated it in collaborative filtering experiments. Compared to state-of-the-art hashing-based baselines, we obtained effectiveness improvements of up to +7\% in NDCG and +14\% in MRR, across 4 widely used datasets. These gains come at no additional cost in storage or recommendation time, as the projected Hamming distance has the same extremely fast computation time as the Hamming distance. Compared to the Hamming distance, we further find that model optimization using the projected Hamming dissimilarity significantly improves the convergence rate, thus speeding up model training.

In future work, we plan to investigate the projected Hamming dissimilarity, and possible adaptions of it, in symmetric retrieval settings consisting of item-item similarity search, as opposed to asymmetric user-item search explored in this work. One such example is document similarity search, which in the hashing setting is known as Semantic Hashing \cite{salakhutdinov2009semantic}, where current work has focused on using the Hamming distance for measuring document similarities \cite{shen2018nash,hansensemhash2019,Chaidaroon:2017:VDS:3077136.3080816,dong-etal-2019-document,hansen2020PairRec,hansen2021multiIndex}. 

\newpage
\bibliographystyle{ACM-Reference-Format}
\balance
\bibliography{main.bib}
\end{document}